\documentclass[aps,twocolumn,nopacs,amsmath,amssymb]{revtex4}
\usepackage{graphicx}
\usepackage[british,UKenglish,USenglish,english,american]{babel}
\usepackage{bm}
\usepackage{amsmath}
\usepackage{amssymb}
\usepackage{float}
\usepackage{xcolor,graphicx}
\usepackage{mathrsfs} % so we can use math script font for Hamiltonian symbol
\begin{document}
	\title{Complete Magneto-Optic Modulation of Lateral and Angular Shifts in Spin-Orbit Coupled Members of the Graphene Family}
	\author{Muzamil Shah}
	\author{Muhammad Sabieh Anwar}
	\email{sabieh@lums.edu.pk}
	\affiliation{Laboratory for Quantum Technologies, Department of Physics, Syed Babar Ali School of Science and Engineering, Lahore University of Management Sciences (LUMS), Opposite Sector U, D.H.A., Lahore 54792, Pakistan}
	\date{\today}
	
	\begin{abstract}
		The intrinsic spin-orbit coupling in the 2D staggered monolayer semiconductors is very large as compared to graphene. The large spin orbit interaction in these materials leads to the opening of a gap in the energy spectrum and spin-splitting of the bands in each valley. In this paper, we theoretically investigate the mechanical steering of beams from these spin-orbit rich, staggered 2D materials. Mechanical steering results in noticeable deviations of the reflected and transmitted ray profiles as predicted from classical laws of optics. These effects are generally called the Goos-Hanchen (GH) and Imbert-Fedorov shifts. We find that electric and magnetic field modulated giant spatial and angular GH shifts can be achieved in these materials for incident angles in the vicinity of Brewster angle in terahertz regime. We also determine the dependence of beam shifts on the chemical potential and find that the Brewster angle and the sign of GH shift can be controlled by varying the chemical potential. This allows the possibility of realizing spin and valley dependent optical effects that can be useful readout markers for experiments in quantum information processing, biosensing and valleytronics, employed in the terahertz regime.
	\end{abstract}
	\maketitle
	
\section{Introduction}
	Monolayer graphene is a thoroughly investigated material and boasts exotic applications due to its fundamentally unique physical electronic and optical properties \cite{Novoselov2004}. For example graphene possesses a gapless Dirac-type band structure \cite{Tombros2007}, high carrier mobilities and universal broadband optical conductivities \cite{Koppens2011}. Recently, staggered 2D semiconductors materials \cite{Castellanos-Gomez2017,Zhao2016} have also attracted intensive attention due to their unusual topological properties \cite{Zhao2016,Grazianetti2016}. These materials include silicene \cite{Molle2017}, germanene \cite{Vogt2012}, stanene \cite{Davila2014}, and plumbene \cite{Saxena2016}. Graphene in general shares some common properties with these materials. For example, these 2D materials have stable honeycomb lattice structures \cite{Cahangirov2009,Cai2015} and are therefore called the graphene family or graphene's siblings. Second, due to larger ionic size of silicon, germanium and tin atoms, they have buckled structures and therefore possess large band gaps. Third, the strong spin-orbit interaction (SOI) is responsible for endowing mass to the Dirac fermions. The strengths of this interaction in silicene \cite{Ezawa2012,Liu2011}, germanene \cite{Vogt2012} and tinene \cite{Xu2013}, are of the order of 1.55\textendash7.9, 24\textendash93, and 100 meV respectively. Further, a static external electric field can also tune the band gap for each spin and valley rendering the Dirac mass controllable in both the $K$ and $K'$ valleys. Under the influence of an electric field, the staggered materials are also known to exhibit different topological phase transitions \cite{Ezawa2013}. Last, analogous to electronics and spintronics, valleytronics provides another degree of freedom (DOF) in these systems, the valley pseudospin, which can be used to store and carry information with implications for quantum information processing \cite{Schaibley2016,Xiao2007}. These materials therefore provide an accessible playground for valleytronics and the possibility to realize novel tuneable magneto-optic (MO) devices \cite{Xiao2012,Zeng2012}.
	
	Goos-H\"{a}nchen (GH) and Imbert-Fedorov (IF) shifts are special beam shift phenomena which occur when a finite extent polarized beam of light is incident on the interface of two dielectric media and the reflected beam undergoes longitudinal (parallel to the plane) and transverse (perpendicular to the plane) shifts, respectively \cite{Goos1947,Imbert1972,Fedorov2013}. These effects are shown in Fig.~\ref{fig1}. During the last decade, the GH shift has been extensively studied in many different systems, such as photonics \cite{He2006,Soboleva2012}, plasmonics \cite{Yin2004,Yin2006}, chiral materials \cite{Huang2017}, metamaterials \cite{Longhi2011} and quantum systems \cite{Lee2014}. The potential applications of GH shift are in biosensors \cite{Yin2006}, optical measurement and optical heterodyne sensors \cite{Hashimoto1989}. For example, controlling the Fermi energy in THz region, L. Jiang \emph{et. al} theoretically studied electrically tunable GH shifts of monolayer graphene \cite{Fan2016}. The magnetic field and Fermi energy modulated giant quantized GH effect on the surface of graphene in the quantum Hall regime was recently predicted \cite{Wu2017}. T. Tang \emph{et. al} proposed an experimental scheme based on a prism-graphene coupling structure for MO tunable GH effect sensing \cite{Tang2014,Tang12018}. Recently the GH shift on the surface of silicene \cite{Wu2018}, TMDCs \cite{Das2018} and Weyl semimetals \cite{Ye2019} has also been investigated. From an application perspective, Farmani \emph{et. al} designed a bidirectional and tunable graphene plasmonic switch in a modified Kretchmann configuration in THz range \cite{Farmani2018}. In the THz to mid-infrared range, the observation of GH effect in graphene may find significant new and interesting applications in extremely sensitive optical sensors \cite{Li2013,Li2014}. Many works are devoted to study the electronic analog of the valley and spin polarized GH shift in silicene and gapped graphene structures \cite{Azarova2017, Wu2011}.

     The novelty of this current work lies in obtaining large spin and valley resolved GH shifts simultaneously in staggered 2D materials in the presence of applied electric and magnetic fields, a scenario not comprehensively covered in previous works. The present study details the impact of magnetic field, chemical potential, incident frequency, electric field and incident angle modulation of the valley- and spin-polarized GH shifts in staggered 2D semiconductors materials.

    Furthermore, the Brewster effect is a fundamental electromagnetic and optical phenomenon in which $p$ polarized light experiences zero reflection \cite{Brewster1815,Lakhtakia1989}. Controlling the Brewster angle is extremely important in broadband devices such as the solid-state modulator \cite{Chen2018}. We also show that the Brewester effect in the silicenic atomic layer is strongly influenced by magnetic fields. Last, we find that angular and spatial GH shifts can be significantly enhanced by tuning the chemical potential, incident beam frequency and the electric field which can render the system in various regimes such as the topological insulator (TI), valley-spin polarized metal (VSPM) and band insulator (BI).
	
	\subsection{Model and Theory}
	\begin{figure*}[!htb]
		\centering
		\begin{tabular}{cc}
			\includegraphics[width=0.5\linewidth]{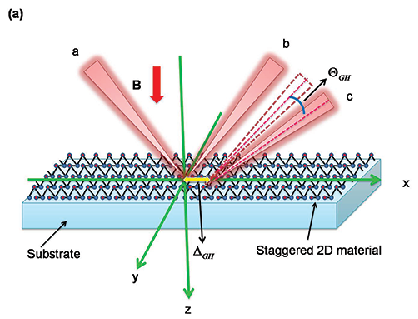}
				\includegraphics[width=0.5\linewidth]{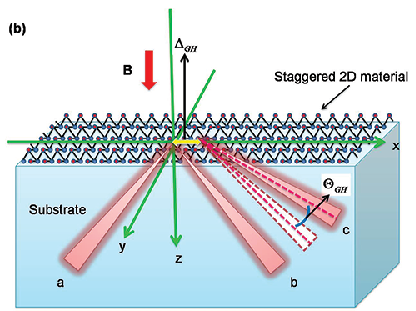}
		\end{tabular}
		\caption{Schematic representation of the beam reflection at a
			2D staggered material-substrate interface in the presence of an external electric and magnetic fields are shown for partial reflection (PR) and total internal reflection (TIR) conditions. The spatial and angular GH shifts for (a) PR and for (b) TIR. The incident, classically pridected and reflected beams are denoted by a, b and c respectively.}
		\label{fig1}
	\end{figure*}
	We consider a well-collimated, monochromatic Gaussian beam of frequency $\Omega$ with a finite beam width impinging from a medium of refractive index $n_{1}$ onto a staggered 2D semiconductor-substrate system as shown in Fig.~\ref{fig1}. The incident angle is $\theta_{1}$ at the interface with a medium of refractive index $n_{2}$. Consider that the $z$ axis of the laboratory Cartesian frame $(x,y,z)$ is normal to the air-staggered 2D monolayer-substrate interface (at $z=0$). A static magnetic field may be applied perpendicularly to the monolayer-substrate system in the $z$ direction. The wave vectors are $k_1$ and $k_2$ where $k_{n}=\Omega\sqrt{\mu_{n}\varepsilon_{n}}$, $Z_n=Z_0\sqrt{\mu_{n}/\varepsilon_{n}}$ and $Z_0=\sqrt{\mu_{0}/\varepsilon_{0}}$, where $\mu_{0}$ and $\varepsilon_{0}$ are the vacuum permeability and permittivity respectively and $n=$1, 2 denote the incident and refracted media.
	The low-energy effective Dirac Hamiltonian for the staggered graphene system can be written as \cite{Tabert2013}
	\begin{equation}\label{a11}
	\hat{H}_{\xi \sigma}=\hbar v_{F}(\xi k_x \hat{\tau}_{x}+k_y \hat{\tau}_{y})-\frac{1}{2}\xi \sigma \Delta_{so}\hat{\tau}_{z}+\frac{1}{2}\Delta_{z}\hat{\tau}_{z}\cdot
	\end{equation}
	Here, $\xi=\pm 1$ denotes the $K$ or $K'$ valley, $v_{F}$ is the Fermi velocity, $\Delta_{so}$ is the intrinsic spin-orbit coupling and $\Delta_z=aE_z$ is the staggered electric potential. The vector operators $\vec{\sigma}=(\hat{\sigma}_{x},\hat{\sigma}_{y},\hat{\sigma}_{z})$ and $\vec{\tau}=(\hat{\tau}_{x},\hat{\tau}_{y},\hat{\tau}_{z})$ represent Pauli matrices of the true and pseudo spin lattice DOF respectively. For Landau level (LL) quantization imposed by an external magnetic field $B$, we introduce the Landau gauge so that the magnetic vector potential $A = (-yB, 0,0)$, and the energy spectrum take the form,
	\begin{equation}\label{a12}
	E(\xi,\sigma, n,t)=\begin{cases}
	t\sqrt{2 v_{F}^{2} \hbar e B |n|+\Delta_{\xi\sigma}^{2}}, & \text{if $n\neq 0$}.\\
	-\xi \Delta_{\xi\sigma}, & \text{if $n=0$}.
	\end{cases}
	\end{equation}
	Here, $t=\textrm{sgn(n)}$ denotes the electron/hole band, $\Delta_{\xi\sigma}=-\frac{1}{2}\xi \sigma \Delta_{so}+\frac{1}{2}\Delta_{z}$ and $n$ is an integer, the quantum number denoting Landau quantization.	The Fresnel reflection coefficients of the staggered 2D monolayer-substrate system in the presence of magnetic field have been spelled out in several previous works and can be expressed as \cite{Kort-Kamp2015,Wu2017}
	\begin{eqnarray}\label{b1}
	r_{pp} &=& \frac{\alpha_{+}^{T}\alpha_{-}^{L}+\beta}{\alpha_{+}^{T}\alpha_{+}^{L}+\beta}, \\
	\label{b2}
	r_{ss} &=& -\bigg(\frac{\alpha_{-}^{T}\alpha_{+}^{L}+\beta}{\alpha_{+}^{T}\alpha_{+}^{L}+\beta}\bigg), \\
	\label{b3}
	r_{sp}=r_{ps}&=&\frac{-2Z_{0}^{2}\mu_{0}\mu_{1}\mu_{2}k_{1z}k_{2z}(\sigma_{xy}^{antisym}+\sigma_{xy}^{sym})}{Z_{1}(\alpha_{+}^{T}\alpha_{+}^{L}+\beta)},
	\end{eqnarray}
	where,
	\begin{eqnarray}\label{b4}
	\alpha_{\pm}^{L} &=& (k_{1z}\varepsilon_{2}\pm k_{2z}\varepsilon_{1}+k_{1z}k_{2z}\sigma_{L}/(\varepsilon_{0}\Omega)), \\
	\label{b5}
	\alpha_{\pm}^{T} &=& (k_{2z}\mu_{1}\pm k_{1z}\mu_{2}+\mu_{0}\mu_{1}\mu_{2}\sigma_{T}\Omega), \\
	\label{b6}
	\beta &=& Z_{0}^{2}\mu_{1}\mu_{2}k_{1z}k_{2z}[(\sigma_{xy}^{antisym})^2-(\sigma_{xy}^{sym})^{2}] \cdot
	\end{eqnarray}
	Here, $k_{1z}=k_{1}\cos(\theta_{1})$ and $k_{2z}=k_{2}\cos(\theta_{2})$. The conductivities $\sigma_{L}(\sigma_{T})$ are the longitudinal (transverse) components. In fact, the Hall conductivity $\sigma_{xy}$ has symmetric $\sigma_{xy}^{sym}$ and asymmetric $\sigma_{xy}^{antisym}$ parts. Here, we are interested only in the antisymmetric part of the conductivity since the symmetric $\sigma_{xy}^{sym}$=0. The corresponding real and imaginary parts of the longitudinal and transverse magneto-optical conductivities of the graphene family computed at $T=0$ K are given by
	\begin{widetext}
		\begin{multline}\label{a6}
		\frac{\textrm{${\textrm{Re} \above 0pt \textrm{Im}}$}\big\} \big(\sigma_{xx}(\Omega)\big)}{\sigma_{0}}=\frac{2 v_{F}^{2}\hbar e B}{\pi}\sum_{\xi, \sigma}\sum_{m,n}\frac{\Theta(E_{n}-\mu_{F})-\Theta(E_{m}-\mu_{F})}{E_{n}-E_{m}}\\\times\bigg[(A_{m}B_{n})^{2}
		\delta_{|m|-\xi,|n|}+(B_{m}A_{n})^{2}\delta_{|m|+\xi,|n|}\bigg]\bigg\{\textrm{${\textrm{F} \above 0pt \textrm{G}}$},
		\end{multline}
		where, $\sigma_{0}=e^2/4\hbar$, $F=\Gamma/\bigg((\hbar\Omega-(E_n-E_m))^{2}+\Gamma^{2}\bigg)$ and $G=\big(\hbar\Omega-(E_n-E_m)\big)/\bigg((\hbar\Omega-(E_n-E_m))^{2}+\Gamma^{2}\bigg)$. In these expressions, the Kronecker deltas ensure the rules for electric dipole transitions between the LL's are satisfied. The absolute values of $m$ and $n$ inside the Kronecker deltas generalizes the case in Ref. \cite{Tabert2013} to both positive and negative chemical potentials $\mu_{F}$, \emph{i.e.} n and p-type 2D materials.
		The Heaviside functions $\Theta(E_{n}-\mu_{F})$ ensure that transitions across the Fermi level are possible, hence they effectively account for the so called Pauli blocking \cite{Grigorenko2012}. Similarly, the real and imaginary parts of the transverse conductivity are
		\begin{multline}\label{a7}
		\frac{\textrm{${\textrm{Re} \above 0pt \textrm{Im}}$}\big\} \big(\sigma_{xy}(\Omega)\big)}{\sigma_{0}}=\frac{2 v_{F}^{2}\hbar e B}{\pi}\sum_{\xi, \sigma}\sum_{m,n}\xi\frac{\Theta(E_{n}-\mu_{F})-\Theta(E_{m}-\mu_{F})}{E_{n}-E_{m}}\\\times\bigg[(A_{m}B_{n})^{2}
		\delta_{|m|-\xi,|n|}-(B_{m}A_{n})^{2}\delta_{|m|+\xi,|n|}\bigg]\bigg\{\textrm{${\textrm{-G} \above 0pt \textrm{F}}$}\cdot
		\end{multline}
	\end{widetext}
	In the limit $\Delta_{so}=\Delta_{z}=0$, we recover graphene's Hall conductivity \cite{Gusynin2005}, as expected.
	
	The $p$ polarized angular and spatial GH of the reflected light from the staggered 2D monolayer-substrate system can be written as \cite{Wu2017}
	\begin{equation}\label{d2}
	\Theta_{GH}=-\frac{2(R_{pp}^{2}\rho_{pp}+R_{ps}^{2}\rho_{ps})}{2k_{1}(R_{ps}^{2}+R_{pp}^{2})\Lambda_{R}+\chi_{pp}+\chi_{ps}},
	\end{equation}
	\begin{equation}\label{d3}
	\Delta_{GH}=\frac{2(R_{pp}^{2}\varphi_{pp}+R_{ps}^{2}\varphi_{ps})\Lambda_{R}}{2k_{1}(R_{ps}^{2}+R_{pp}^{2})\Lambda_{R}+\chi_{pp}+\chi_{ps}}.
	\end{equation}
	where, $\Lambda_{R}=\pi w_{0}^2/\lambda$ is the Rayleigh range,  $r_\lambda=R_\lambda\exp(i\phi_{\lambda})$, $\lambda \in (pp,ps,ss,sp)$, $\rho_\lambda=\textrm{Re}(\partial \ln r_{\lambda}/\partial\theta_{1})$, $\varphi_\lambda=\textrm{Im}(\partial \ln r_{\lambda}/\partial\theta_{1})$ $\chi_\lambda=R_{\lambda}^{2}(\varphi_{\lambda}^{2}+\rho_{\lambda}^{2})$, $R_\lambda$ is the amplitude and $\phi_{\lambda}$ is the phase of reflection coefficients.
	
\section{Results and Discussion}
\subsection{Magnetic field modulated GH under PR and TIR conditions}
\begin{figure*}[!t]
	\centering
	\begin{tabular}{cc}
		\includegraphics[width=0.9\linewidth]{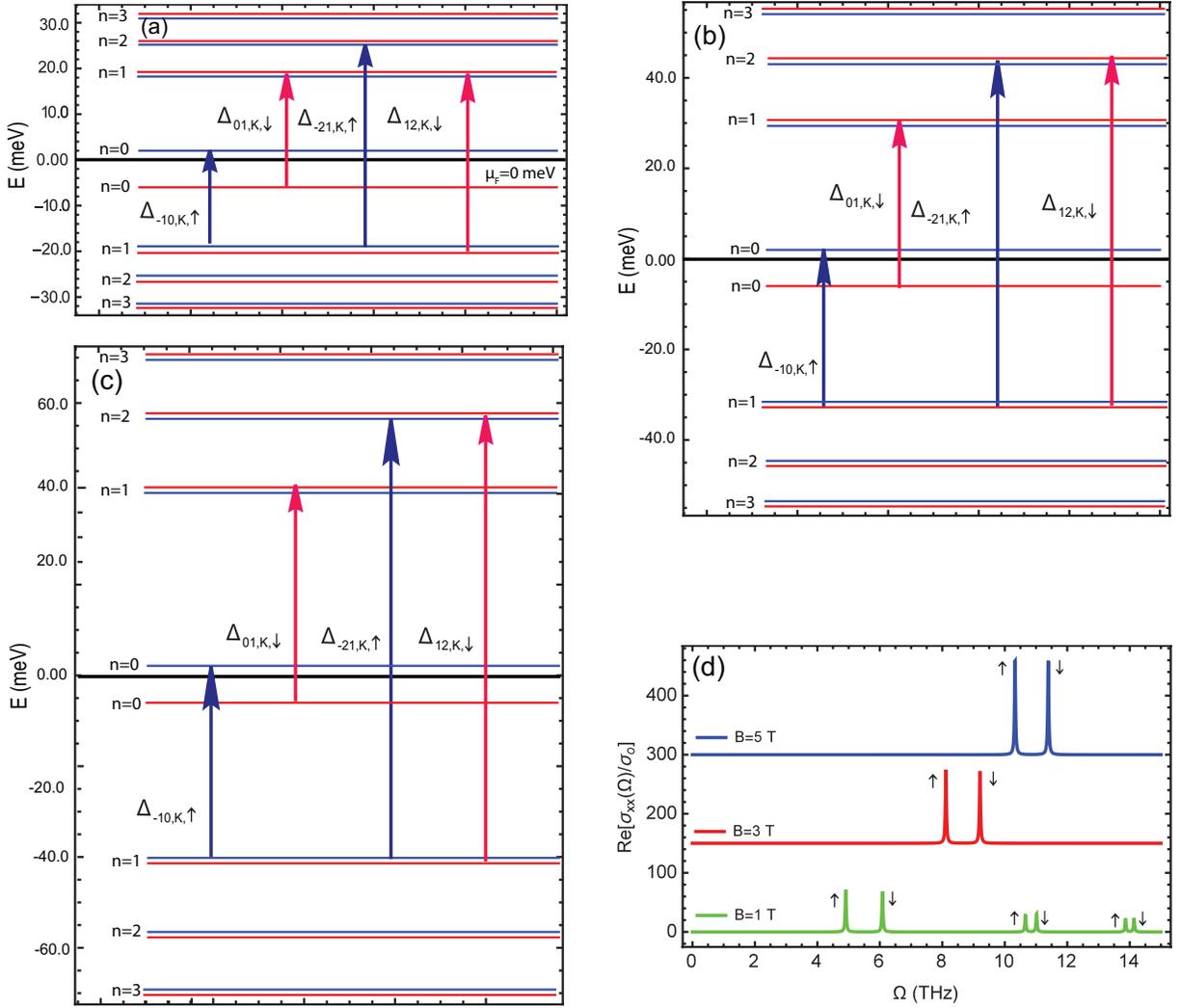}
	\end{tabular}
	\caption{Schematic representation of the allowed transitions between LL’s for three
		different magnetic fields in the $K$ valley at (a) 1 T (b) 2 T (c) 3 T. Blue lines represent Landau levels for spin up ($\sigma=\uparrow$) and red lines represent Landau level for spin down ($\sigma=\downarrow$). The same color scheme applies for the Landau levels transitions. (d) Longitudinal conductivity as a function of photon frequency. The parameters used are $\Delta_{z}=0.5\Delta_{so}$ and chemical potential $\mu_{F}=0$.}
	\label{fig2}
\end{figure*}
We quickly recount the effect of the magnetic field on the energy level structure and subsequently the MO response. Schematic diagrams showing the allowed transitions between Landau levels (LL’s) for three different magnetic field ($B=$1, 3 and 5 T) all in the topological insulator regime $(\Delta_{z} < \Delta_{so})$ and in the $K$ valley are shown in Fig.~\ref{fig2}(a)\textendash(c). The transition energy is determined from the energy difference between LL's obeying certain selection rules namely $|n|-|m|=\pm1$ and the conservation of real spin implying that transitions between $\sigma=+1$ and $-1$ levels are spin forbidden. The excitation energies corresponding to the different transitions, $E_{m,K (K'),\uparrow(\downarrow)}\rightarrow E_{n,K (K'),\uparrow(\downarrow)}$ are labelled as $\Delta_{mn,K (K'),\uparrow(\downarrow)}$. Blue lines represent Landau levels for spin up ($\sigma=\uparrow$) and red lines represent Landau levels for spin down ($\sigma=\downarrow$). In each of the depicted transitions, one of the participating levels is an $n$=0 level. For example for $B=1$ T, the first and second magneto-excitation energies correspond to the $\Delta_{-10,K, \uparrow}$ and $\Delta_{01,K, \downarrow}$ for spin up and spin down respectively and are calculated at 20.3 meV (4.9 THz) and 25.1 meV (6.1 THz) respectively. For higher $n$, the LL's are closely spaced, we ignore these transitions for the time being. Similarly for $B=3$ and 5 T, the allowed transitions between LL's are shown in Fig.~\ref{fig2}(b) and (c). In fact, Table I summarizes magneto-excitation frequencies within the $n=-1,0,1$ manifold for the magnetic fields considered. In Fig.~\ref{fig2}(d), we have shown the variation of the longitudinal conductivity as a function of photon frequency for three different magnetic fields also shown exclusively for the TI regime. We can see resonant peaks when the incident $\hbar\Omega$ hits the magneto-excitation energy. As we increase the strength of the applied magnetic field, the MO excitations shift towards higher frequencies. For practicality and simplicity, we will restrict ourselves to the lowest magneto-excitation transition frequency, originating only from the $\Delta_{-10,K, \uparrow}$ transition while investigating the GH shift in this article, unless otherwise specified.

\begin{figure*}[!ht]
	\centering
	\begin{tabular}{cc}
		\includegraphics[width=0.9\linewidth]{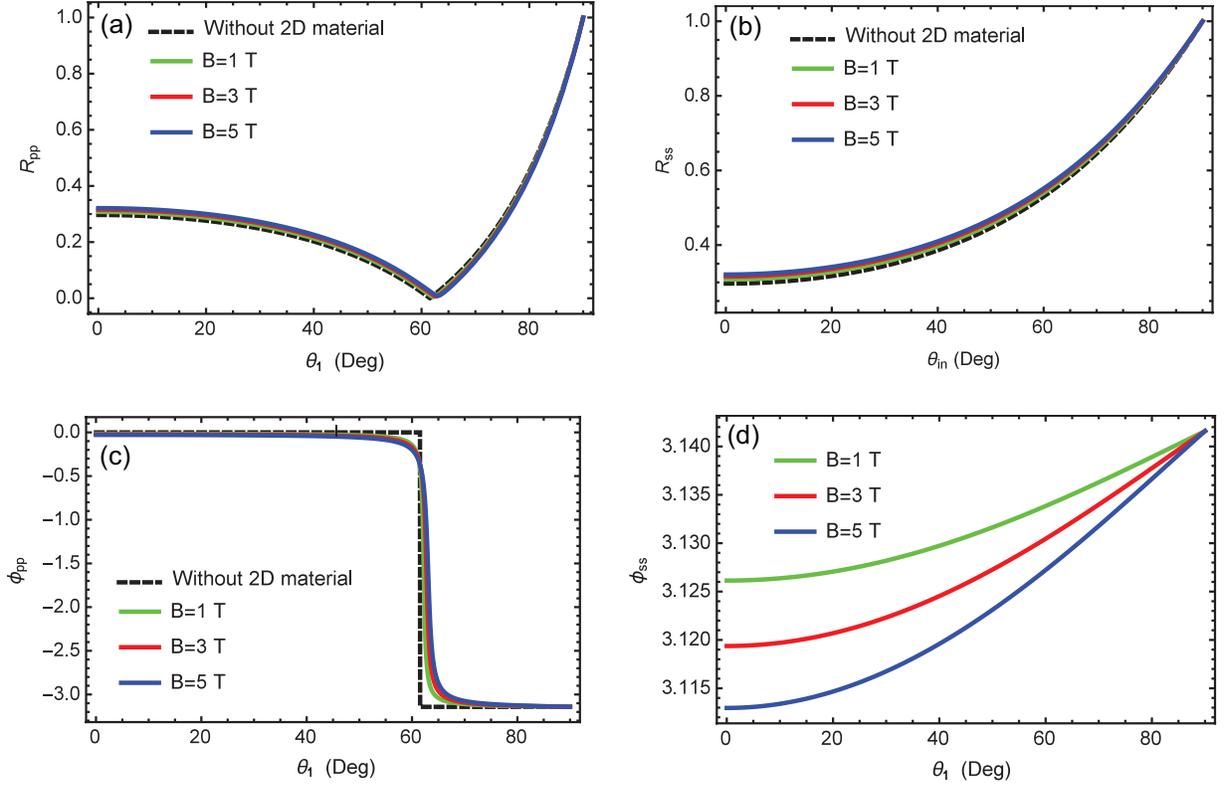}&
	\end{tabular}
	\caption{Modulus and phase of the $s$ and $p$ polarized reflection coefficients for 2D staggered graphene-substrate system as a function of incident angle for different magnetic fields in $K$ valley for PR: (a) $R_{ss}$, (b)$R_{pp}$, (c) $\phi_{pp}$ and (d) $\phi_{ss}$. The parameters used are $\Gamma=0.2\Delta_{so}$, refractive index $n_{2}=1.84$ and chemical potential $\mu_{F}=0$.}
	\label{fig3}
\end{figure*}

\begin{figure*}[!t]
	\centering
	\begin{tabular}{cc}
		\includegraphics[width=0.90\linewidth]{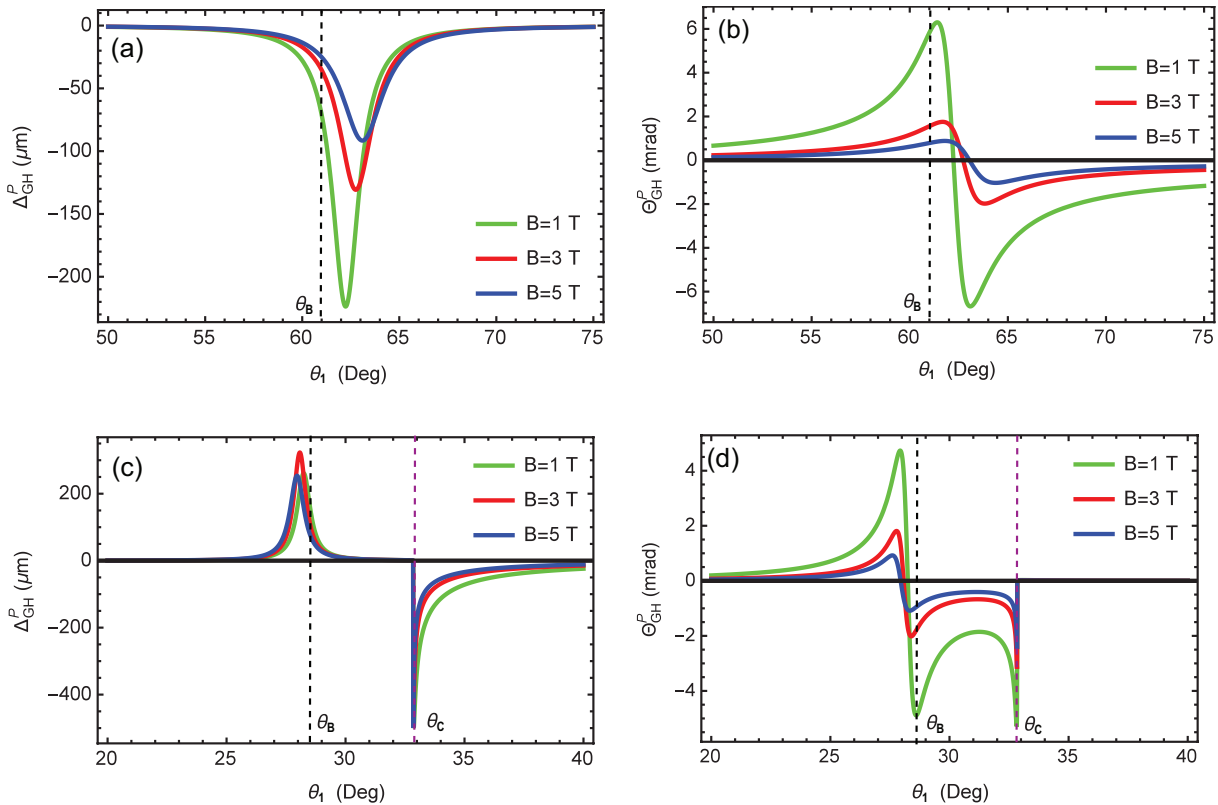}
	\end{tabular}
	\caption{The $p$ polarized spatial and angular GH shifts for charge neutral staggered graphene-substrate system as a function of incident angle for different magnetic fields in the $K$ valley in the TI regime for PR and TIR. (a) The $p$ polarized spatial GH shifts for PR, (b) the $p$ polarized angular GH shifts for PR, (c) the $p$ polarized spatial GH shifts for TIR and (d) the $p$ polarized angular GH shifts for TIR. The dashed lines represents the values of $\theta_{B}$ and $\theta_{C}$ for the native dielectric substrate. The parameters used are identical across all figures, unless stated otherwise. (A color version of this figure can be viewed online).}
	\label{fig4}
\end{figure*}

With knowledge of the LL transitions and magneto optical conductivities of the 2D materials for different magnetic fields at hand, we are now capable to studying the spatial and angular GH shifts. We first discuss the reflectivity and phases of the reflected $s$ and $p$ polarized waves for incident light. Clearly, if the impinging Gaussian beam frequencies are smaller than all Dirac gaps, then no electrons can be excited from the valence to the conduction band. On the other hand, a resonant Gaussian wave will excite a electron-hole pair. As the resonance condition is spin-dependent, so spins of only one kind (say up spin) electrons will be excited. The resonant excitation energies (frequencies) for the $\Delta_{-10,K, \uparrow}$ transition are 20.3 meV (4.9 THz) at $B = 1$ T; 33.5 meV (8.1 THz) at $B = 3$ T and 42.6 meV (10.3 THz) at $B = 5$ T. Fig.~\ref{fig3} illustrates the magnetic field modulated reflectivity ($R_{ss}$ and $R_{pp}$) and their phases ($\phi_{ss}$ and $\phi_{pp}$) as a function of incident angle in the TI regime. In Fig.~\ref{fig3}(a) and (b), the moduli of $R_{pp}$ and $R_{ss}$ are shown for various magnetic fields indicating that for $p$ polarized incident light, $R_{pp}$ achieves a minimum value at a certain $\theta_{1}$ and rises again. This is called the pseudo-Brewster angle $\theta_{B}=\tan^{-1}(n_{2}/n_{1})$, whereas $R_{ss}$ increases smoothly as the angle of incidence is increased.  At $\theta_{B}$, the magnitude of reflection coefficient $R_{pp}$ intensity for the uncoated surface (without 2D material) reaches zero, as shown in Fig.~\ref{fig3}(a) by dashed line. For the 2D staggered atomic layer, $R_{pp}$ at the Brewster angle $\theta_{B}$ is non-zero, because the MO conductivity of the graphene-family is complex. The phase $\phi_{pp}$ shows transition from $0$ to $-\pi$ in the vicinity of $\theta_{B}$ for different magnetic fields. A similar variation has been reported in 2D-TMDC \cite{Das2018}. The phase $\phi_{ss}$ shows an increasing trend with $\theta_{1}$ in the TI regime, as shown in Fig.~\ref{fig3}(d). Note that we haven't shown results in the VSPM state and BI regime, for which $R_{ss}$ and $R_{pp}$ don't appreciably change for magnetic fields. In the reminder of this section, we will only discuss the magnetically induced spatial and angular shifts for the interesting case of $p$ polarized light. The shifts for the charge neutral graphene-family ($\mu_{F}=0$ meV) are plotted as a function of incident angle for different magnetic fields in the TI regime in Fig.~\ref{fig4}(a), whereas the angular GH shifts are represented in Fig.~\ref{fig4}(b). Since $\partial\phi_{p}/\partial\theta_{1}\ne 0$ and is accentuated in the vicinity of $\theta_{B}$, the spatial shifts are dominant in the vicinity of the pseudo-Brewster angle.

	\begin{table*}[!htbp]
	\textbf{\caption{Table of allowed transitions in $K$ valley in the $n=-1,0,1$ subspace, for different magnetic fields in the TI regime with $\Delta_{so}=0.5\Delta_{z}$.}}
	\centering
	\begin{tabular}{lllll}
		\hline
		$\Delta_{mn, K(K'),\uparrow(\downarrow)}$  &~~~~~~ $B$ (T)  &~~~~~~~~~~~~Frequency (THz) \\ \hline
		$\Delta_{-10, K, \uparrow}$ &~~~~~~~~~1 &~~~~~~~~~~~~~~~~~~~~~$4.9$  \\
		$\Delta_{01, K, \downarrow}$ &~~~~~~~~~1 &~~~~~~~~~~~~~~~~~~~~~$6.1$  \\
		$\Delta_{-10, K, \uparrow}$ &~~~~~~~~~3 &~~~~~~~~~~~~~~~~~~~~~$8.1$  \\
		$\Delta_{01, K, \downarrow}$ &~~~~~~~~~3 &~~~~~~~~~~~~~~~~~~~~~$9.2$  \\
		$\Delta_{-10, K, \uparrow}$ &~~~~~~~~~5 &~~~~~~~~~~~~~~~~~~~~~$10.3$  \\
		$\Delta_{01, K, \downarrow}$ &~~~~~~~~~5 &~~~~~~~~~~~~~~~~~~~~~$11.4$  \\
	\end{tabular}
\end{table*}

To better visualize the dependence of the GH shifts with respect to the magnetic field consider the $p$ polarized spatial GH shifts in Fig.~\ref{fig4}(a), for the three different magnetic fields. We observe giant negative spatial beam shifts in the TI regime. If we look at the $\phi_{pp}$ spectra in Fig.~\ref{fig3}(c) together with these shifts, we observe that $\phi_{pp}$ shows a change of $-\pi$ resulting in negative spatial GH shifts. The magnitude of spatial GH shifts are larger for smaller magnetic fields and smaller for large magnetic fields values; however, the extent of the spatial GH shifts for $p$ polarization on a 2D staggered monolayer graphene-substrate system is larger compared to the case of the graphene-coated surface \cite{Grosche2015}. It is also worth noting that the Brewster angle is also sensitive to the applied magnetic field. From the results shown in Fig.~\ref{fig4}(a), we observe that the Brewster angle of the spatial GH shifts can be tuned from $62^{\circ}$ to $63^{\circ}$ by modulating magnetic field for the three proposed magnetic fields.

Similarly in Fig.~\ref{fig4}(b), we also plot the angular GH shifts as a function of the incident angle. The angular GH shift $\Theta_{GH}^{p}$ is positive and gradually increases with incident angles, but as the incident angle reaches $\theta_{B}$, $\Theta_{GH}^{p}$ decreases rapidly and results in a negative GH shift. By tuning the angle of incidence of the $p$ polarized THz beam and the applied magnetic field, one can therefore control the polarity of the angular GH shift as well as its amplitude. Similar to the spatial counterpart, the amplitude of the angular GH decreases for increasing magnetic fields and this finding is in good agreement with purely graphene-coated surfaces \cite{Grosche2015}. Once again, the tuning of the Brewster angle $\theta_{B}$ with magnetic field is also clearly observable.

Similarly we plot the $p$ polarized spatial and angular GH shifts under the total internally reflected (TIR) condition in Figs~\ref{fig4}(c) and (d) respectively. Notable deviations from external reflection and certain additional features are vividly observable here. From Fig.~\ref{fig4}(c) we observe that for magnetic fields of 1, 3 and 5 T, we have peaks in the vicinity of $\theta_{B}$ and sharply precipiced dips at the critical angle $\theta_{C}=\sin^{-1}(n_{2}/n_{1})$, in the TI regime. Once again, the amplitude of the spatial GH shift is attenuated for larger magnetic fields. Similarly the magnetic field modulated $p$ polarized angular GH shifts for TIR geometry are depicted in Fig~\ref{fig4}(d) wherein we expect unusual behavior of the GH shifts in the vicinity of both the Brewster and the critical angle. The angular GH shift $\Theta_{GH}^{p}$ is positive when the incidence angle is smaller than the Brewster angle and negative beyond. Similarly, just prior to the critical angle, the angular shift acquires a large negative spike and diminishes immediately after. Furthermore, near $\theta_{B}$, the angular GH decreases in size with increasing magnetic field values.

\begin{figure*}[!t]
	\centering
	\begin{tabular}{cc}
		\includegraphics[width=0.90\linewidth]{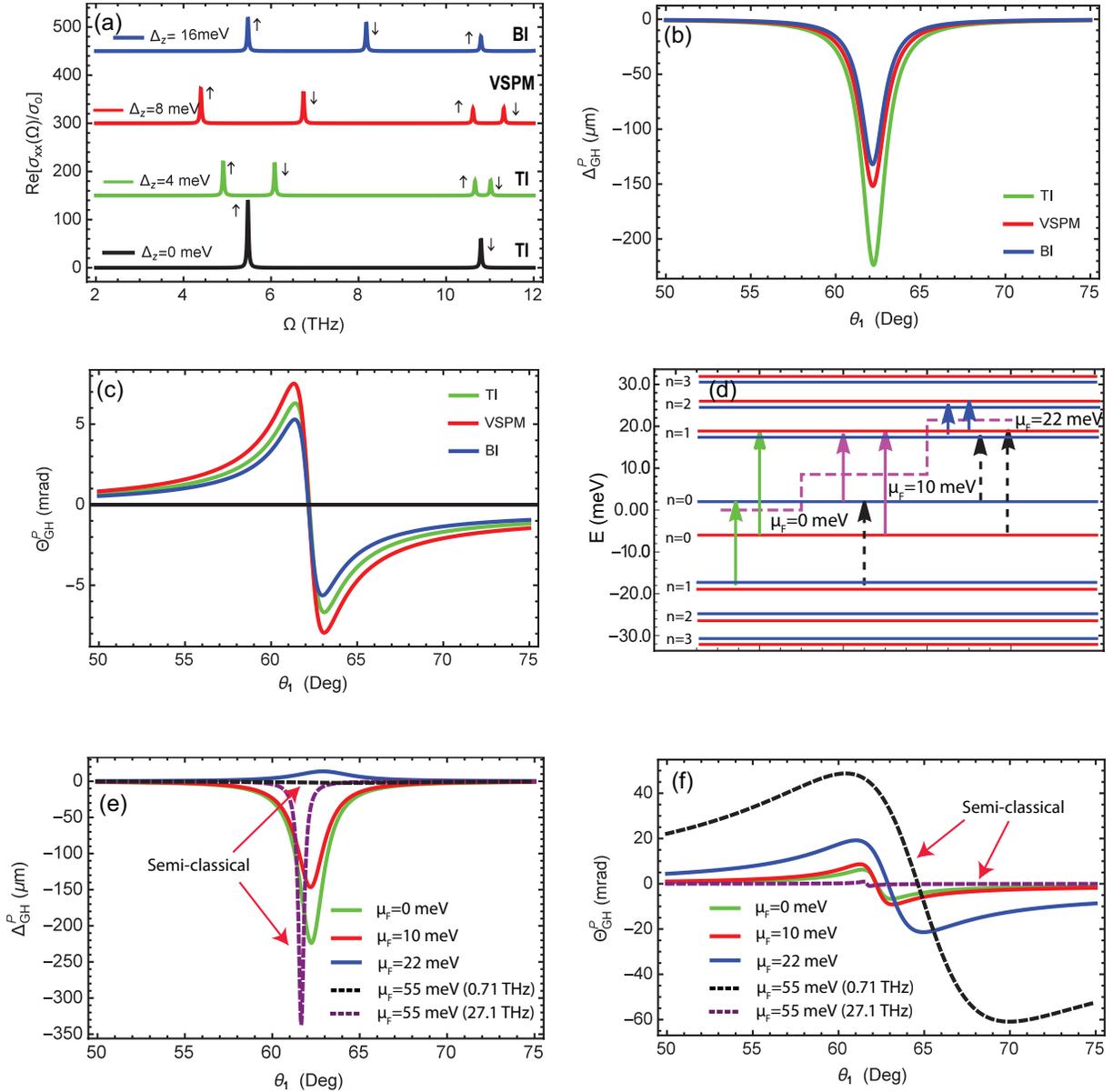}
	\end{tabular}
	\caption{(a) Longitudinal conductivity as a function of incident photon frequency and the $p$ polarized spatial and angular GH shifts for the staggered graphene-substrate system as a function of incident angle for magnetic field $B=1$ T in the $K$ valley in three distinct topological regimes and for four different chemical potentials. The incidence is external PR, while (b) and (c) show the $p$ polarized spatial and angular GH shifts with modulation of the external electric field, for the TI, VSPM and BI at a magnetic field of 1 T. (d) Schematic representation of the allowed transitions between LL’s for three different values of chemical potential $\mu_{F}$=0, 10 and 22 meV, and (e) and (f) are the $p$ polarized spatial and angular GH shifts with modulation of the chemical potential in the TI and classical regimes for a magnetic field of 1 T.}
	\label{fig5}
\end{figure*}

\begin{table*}[!htb]
	\textbf{\caption{Table of allowed transitions in $K$ valley in the $n=-1,0,1$ subspace, for $B=1$ T in three different topological regimes for $\Delta_{so}=8$ meV.}}
	\centering
	\begin{tabular}{lllll}
		\hline
		$\Delta_{mn, K(K'),\uparrow(\downarrow)}$  &~~~~~~ $\Delta_{z}$ (meV) &~~~~~~ Regime &~~~~~~~~~~~~Frequency (THz) \\ \hline
		$\Delta_{-10, K, \uparrow}$ &~~~~~~~~~0 &~~~~~~~~~TI &~~~~~~~~~~~~~~~~~~~~~$5.6$  \\
		$\Delta_{01, K, \downarrow}$ &~~~~~~~~~0 &~~~~~~~~~TI &~~~~~~~~~~~~~~~~~~~~~$5.6$  \\
		$\Delta_{-10, K, \uparrow}$ &~~~~~~~~~4 &~~~~~~~~~TI &~~~~~~~~~~~~~~~~~~~~~$4.9$  \\
		$\Delta_{01, K, \downarrow}$ &~~~~~~~~~4 &~~~~~~~~~TI &~~~~~~~~~~~~~~~~~~~~~$6.1$  \\
		$\Delta_{-10, K, \uparrow}$ &~~~~~~~~~8 &~~~~~~~VSPM &~~~~~~~~~~~~~~~~~~~~~$4.4$  \\
		$\Delta_{01, K, \downarrow}$ &~~~~~~~~~8 &~~~~~~~VSPM &~~~~~~~~~~~~~~~~~~~~~$6.7$  \\
		$\Delta_{-10, K, \uparrow}$ &~~~~~~~~16 &~~~~~~~~~BI &~~~~~~~~~~~~~~~~~~~~~$5.5$  \\
		$\Delta_{01, K, \downarrow}$ &~~~~~~~~16 &~~~~~~~~~BI &~~~~~~~~~~~~~~~~~~~~~$8.2$  \\
	\end{tabular}
	\label{11}
\end{table*}

\subsection{Electric field modulated GH shift}
A static electric field $E_{z}$, controls the electronic band structure of the 2D staggered graphene family by generating a staggered sublattice potential $\Delta_{z}$. An increase in the electric field triggers a well known quantum phase transition occurs from topological insulator to band insulator state \cite{Ezawa2012}. Fig.~\ref{fig5}(a) shows the longitudinal conductivity verses photon energy for various values of $\Delta_{z}$. As we increase the applied electric field $\Delta_{z}$, each interband peak splits into two spin-polarized peaks in the TI regime $(\Delta_{z} < \Delta_{so})$. Concomitantly, due to a redistribution of spectral weight, the intensity of the peaks is reduced for larger fields values. When $\Delta_{z} = \Delta_{so}$, the gap of one of the spin-split bands closes and a new type of metallic phase emerges called the valley-spin polarized metal (VSPM) state. At the VSPM point, the lowest frequency peaks, move apart: the $\Delta_{-10,K, \uparrow}$ peak is red shifted while the $\Delta_{01,K, \downarrow}$ peak is blue shifted. The excitation energies corresponding to the first two peaks at the VSPM point are now 18.2 meV (4.4 THz) and 27.8 meV (6.7 THz) for spin up and down, respectively. Further increasing $\Delta_{z}$ results in re-opening of the gaps and the system transitions from the VSPM to the band insulator (BI) state. In BI regime all interband peaks move to higher energies. The magneto-excitations frequencies are presented in Table II for the first two transitions in the three different regimes. The main role of the electric field is that it controls the band structure and is responsible for spin and valley polarized responses, and therefore, similar to magnetic fields, also controls the magneto-optic excitation energies.

In Figs.~\ref{fig5}(b) and (c), we show the $p$ polarized spatial and angular GH shifts as a function of the incident angle in the TI, VSPM and BI regimes, illustrating selective excitation of spin up carriers in the $K$ valley. From Fig.~\ref{fig5}(b), it is obvious that $p$ polarized incident light produces giant negative spatial GH shifts. For example, in the TI regime when all Dirac gaps are open and at the magneto-excitation frequency of 4.9 THz, a maximum value of $\Delta_{GH}^{p}\approx222$ $\mu$m is attained for scattering rate of $\Gamma=0.2\Delta_{so}$. Further increasing $\Delta_{z}$ results in reopening the lowest energy gaps and the system undergoes a topological phase transition from VSPM state to the BI regime. After band inversion, we observed smaller shifts, $\Delta_{GH}^{p}=130$ $\mu$m. By increasing the electric field, the locations of the Brewster angle does not change but the spatial GH shifts diminish in size.

The $p$ polarized angular GH shift as a function of the incident angle in the TI, VSPM and BI regimes can also be seen in Fig.~\ref{fig5}(c), showing first, the sign inversion of the angular displacement across Brewster's angle and second, illustrating that the impact of the electric field is to suppress the size of the angular shifts.

\subsection{Controlling the magnitude, sign and the position of GH shifts through the  chemical potential}
The THz magneto-optical conductivity is heavily influenced by the chemical potential $\mu_{F}$. We examine the $p$ polarized beam shifts by modulating the chemical potential solely in the $K$ valley. Results for the $K'$ valley are analogous. We observe that the magnitude, sign and position of the Brewster's angle can all be controlled by this external stimulus. For this purpose we choose three different values of chemical potential of $\mu_{F}$=0, 10 and 22 meV, while keeping the magnetic field 1 T positioning the system in the TI regime. In contrast to neutral staggered graphene ($\mu_{F}$=0 meV), we now have interband as well as intraband transitions. For $\mu_{F}$=10 meV the chemical potential lies in between the $n$=0 and $n$=1 LL's and for $\mu_{F}$=22 meV the chemical potential is in between the $n$=1 and $n$=2 LL's. A schematic diagram which helps us to understand the possible LL's transitions for this chemical potential adaptation is shown in Fig.~\ref{fig5}(d). Moving from left to right, we can see the LL's interband transitions for $\mu_{F}=0$ meV are energies $\Delta_{-10,K,\uparrow}=20.3$ meV and $\Delta_{01,K,\downarrow}=25.1$ meV. As $\mu_{F}$ increases to $10$ meV, then according to selection rules, certain transitions become Pauli blocked. For instance, the interband transition $\Delta_{-10,K,\uparrow}$ becomes forbidden which is shown by the dashed upward pointing arrow in the middle. In its lieu, the intra-band transition $\Delta_{01,K,\uparrow}=16.0$ meV emerges whereas for the spin down electron, we still have the allowed interband transition $\Delta_{01,K,\downarrow}=25.1$ meV, shown by a solid arrow. If the chemical potential $\mu_{F}$ is further enhanced to $22$ meV, so that it lies between the $n$=1 and $n$=2 manifolds, then both of the transitions $\Delta_{-10,K,\uparrow}$ and $\Delta_{01,K,\downarrow}$ become Pauli blocked, which are again indicated by the dashed arrows in the rightmost part of Fig.~\ref{fig5}(d). In this case, the excitation energies (frequencies) corresponding to the first two intra-band transitions $\Delta_{12,K,\uparrow}$ and $\Delta_{12,K,\downarrow}$ are 7.5 meV (1.8 THz) and 7.2 meV (1.6 THz), respectively. Table III summarizes these results.

\begin{table*}[!htb]
	\textbf{\caption{Table of allowed transitions in $K$ valley in the $n=-1,0,1$ subspace, for different chemical potentials in the TI regime with $\Delta_{so}=8$ meV.}}
	\centering
	\begin{tabular}{lllll}
		\hline
		$\Delta_{mn, K(K'),\uparrow(\downarrow)}$  &~~~~~~ $\mu_{F}$ (meV)  &~~~~~~  Inter/intraband &~~~~~~~~~~~~Frequency (THz) \\ \hline
		$\Delta_{-10, K, \uparrow}$ &~~~~~~~~~0 &~~~~~~  Inter &~~~~~~~~~~~~~~~~~~~~~$4.9$  \\
		$\Delta_{01, K, \downarrow}$ &~~~~~~~~~0 &~~~~~~  Inter &~~~~~~~~~~~~~~~~~~~~~$6.1$  \\
		$\Delta_{01, K, \uparrow}$ &~~~~~~~~10 &~~~~~~  Intra &~~~~~~~~~~~~~~~~~~~~~$4.0$  \\
		$\Delta_{01, K, \downarrow}$ &~~~~~~~~10 &~~~~~~  Inter &~~~~~~~~~~~~~~~~~~~~~$6.1$  \\
		$\Delta_{12, K, \uparrow}$ &~~~~~~~~22 &~~~~~~  Intra &~~~~~~~~~~~~~~~~~~~~~$1.8$  \\
		$\Delta_{12, K, \downarrow}$ &~~~~~~~~22 &~~~~~~ Intra &~~~~~~~~~~~~~~~~~~~~~$1.6$  \\
	\end{tabular}
\end{table*}

Figs.~\ref{fig5}(e) and (f) depicts the impact of altering the chemical potential (e.g. by chemical doping) on the GH shifts.  Only the $K$ valley is demonstrated for the sake of brevity. For these results, we selected the resonance frequency that excites only the first allowed LL transition for the spin up electrons. The $p$ polarized spatial GH shifts as a function of incident angle for different chemical potentials are shown in Fig.~\ref{fig5}(e). For $\mu_{F}$=0 meV, we have negative spatial GH shifts originating from the $\Delta_{-10,K,\uparrow}$ transition, which is purely interband. When the chemical potential $\mu_{F}$=10 meV, we have the intra-band transition and the magnitude of the $p$ polarized spatial GH shift is larger. As we further increase to $\mu_{F}$=22 meV, we have purely intraband transitions. Here the sign of GH shift switches which is a remarkable demonstration of chemical potential modulated angular GH shifts. Note that an intraband transition is responsible for the giant positive GH shift, precisely $\Delta_{12, K, \uparrow}$ and whose resonant frequency is 1.8 THz. With a further increase of chemical potential, the magnitude of the $p$ polarized spatial GH shift decreases while its full width half maximum increases. Furthermore, we conclude that the pseudo-Brewster angles shifts to larger incidence angles due to the chemical modification. The inversion of sign of the GH shift from negative to positive has significant applications in optoelectronic devices and chemical potential measurement.

\begin{figure*}[!htb]
	\centering
	\begin{tabular}{c}
		\includegraphics[width=0.90\linewidth]{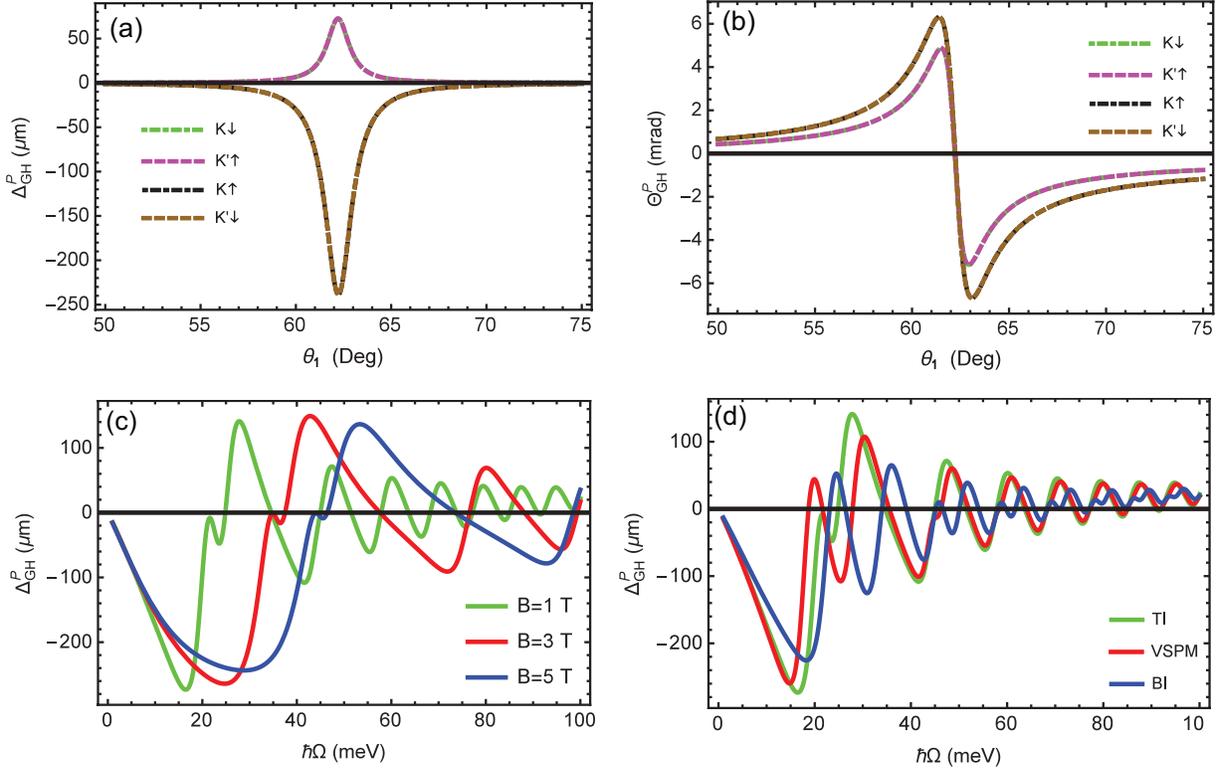}
	\end{tabular}
	\caption{The valley and spin polarized spatial and angular GH shifts for staggered 2D material-substrate system as a function of incident angle for $K$ and $K'$ valleys in the TI regime. (a) The spatial GH shifts for both spins and for both valleys, (b) the $p$ polarized angular GH shifts for both spins and valleys. The spatial and angular GH shifts for staggered 2D material-substrate system as a function of photon energy in the $K$ valley for different magnetic and electric fields. (c) the $p$ polarized spatial and angular GH shifts for three different magnetic fields in the TI regime, (d) the spatial and angular GH shifts for $B=1$ T in three distinct topological regimes. The $p$ polarized response is shown only.}
	\label{fig6}
\end{figure*}

In the semiclassical approximation the LL spacing becomes unimportant, which occur when $|\mu_{F}|\gg|E_{0}|$, or another words when $\mu_{F}$ is high up in the conduction band or deep down in the valance band \cite{Shah2019}. In this regime, the conductivities can be modeled by classical Drude-like peaks. For example we suppose that $\mu_{F}$=55 meV which places the chemical potential between the $n$=9 and $10$ LL's. According to the selection rules, only two transitions $\Delta_{-9~10}$ and $\Delta_{9~10}$ are allowed \cite{Shah2019}. The magneto excitation frequencies for $\Delta_{-9~10}$ and $\Delta_{9~10}$ are 27.1 and 0.71 THz respectively. We plot the spatial and angular GH shifts for this semi-classical scenario in Figs.~\ref{fig5}(e) and (f). We can see in Figs.~\ref{fig5}(e) a giant negative and minuscule spatial GH shift for the incident beam exciting the $\Delta_{-9~10}$ and $\Delta_{9~10}$ transitions respectively. We also observe that the Brewster angle is strongly influenced by the chemical potential and dramatically changes in the classical regime for the two different transitions. Similarly we plot the $p$ polarized angular GH shifts in Fig.~\ref{fig5}(f) for the two mentioned transitions. Remarkably, we get giant angular shift for $\Delta_{9~10}$ transition as compared to a small effect for the $\Delta_{-9~10}$ transition.

\subsection{Spin and valley polarized GH shifts}
In addition to charge and spin degree of freedom, the Dirac electrons in staggered 2D atomic layers possess a valley degree of freedom which acts like a fictitious spin $1/2$ particle. The valley information for our system is actually embodied in the variable $\xi$, which is described in the system Hamiltonian in Eq.~(\ref{a11}). Since the MO excitation energies are spin and valley dependent, one can selectively address the valley pseudospin in these materials to achieve fully spin and valley-polarized GH shifts. For instance, by impinging right or left handed circularly polarized light of the correct frequency, we can selectively excite spin up or spin down electrons in either of the $K$ or $K'$ valleys. Hence we have full freedom to choose spin or valley by tuning the frequency and helicity of the impinging radiation.

In this vein, the valley and spin-dependence of the spatial GH shifts are depicted in Fig~\ref{fig6}(a), indicating positive and negative GH shifts for both $K$ and $K'$ valleys. It is clear that this valley-dependent response is quite interesting. The sign of the lateral GH shift is inverted across the two valleys for the same spin identity. However, the angular shift does \emph{not} undergo sign-inversion across the two valleys. Therefore for experimental verification, the direction of the lateral shift can allow probing of the valley polarization tuning it into a prospective readout modality for the valley qubit in quantum information processing schemes.

\subsection{Incident photon frequency dependence of GH shifts}
As we vary the magnetic field imposed on the 2D atomic layer, the energy and hence the MO excitation energies manifold change. It is therefore instructive, to examine the dependence of the spatial and angular GH shifts on the frequency (energy) of the incident THz beam. In Fig.~\ref{fig6}, we plot the beam shifts as a function of incident photon energy. In Fig.~\ref{fig6}(c) we show the response for three different values of $B=$1, 3 and 5 T, while keeping $\mu_{F}=0$ and $\theta_{1}=61^{\circ}$ (near the Brewster angle) and keeping the electric field fixed to assign the system to the TI regime. Only the $K$ valley spin up polarized response is demonstrated. The beam shifts display an oscillating dependence moving gradually to higher frequencies as the magneto excitation energy is increased. This is shown in Fig.~\ref{fig6}(c). Likewise for a fixed magnetic field ($B=1$ T), as the electric field is changed, rendering the system into various topological regimes, the position as well as the magnitudes of the beam shifts change. From the TI$\rightarrow$VSPM$\rightarrow$BI progression, the GH shifts increase and then decreases. This is shown in Fig.~\ref{fig6}(d).

\section{Conclusion}

In summary, we have systematically investigated the novel spatial and angular mechanical Goos-Hanchen shifts by impinging a Gaussian beam on the surface of a staggered 2D monolayer of the graphene family. The lateral shifts are modulated by electric and magnetic fields and are analyzed in the THz frequency range. By taking into account the fascinating tunable electro-optic and MO properties of these materials, we have studied the effect of chemical potential, spin, valley and incident photon frequency dependence on the silicenic layer, with the particular enhancement of the GH shift in the vicinity of Brewster angle. Furthermore, we have found that the Brewster angle is sensitive to changing magnetic field and chemical potential.

The MO controlled Brewster angle can be used to develop a highly tunable solid-state modulator. More interestingly, the valleys and spins indices can be used for the switching of the GH shift from negative to positive and vice versa.  In conclusion, the GH shift in staggered 2D materials paves the way of realizing spin and valley dependent devices and systems that can be useful optical readout markers for experiments in quantum information processing, biosensing, spintronics and valleytronics in the THz regime. Moreover, the demonstrated electro-optic and MO tunable GH shifts can be used for the detection of electric field and magnetic fields sensing and also for the determination of the doping level of these 2D staggered layers.

\section*{Acknowledgements}
The authors would like to acknowledge financial support from the National Research program for Universities (NRPU), scheme number 10375 funded by the Higher Education Commission of Pakistan.

\end{document}